\documentclass[a4paper,11pt]{article}
\usepackage{aaskaiid}
\usepackage{orcidlink}
\newcommand{\aastar}{AA$^{\ast}$}

\usepackage{xcolor}
\usepackage{hyperref}

\title{Transforming X-ray Binary Astrophysics with SKA+VLBI}
\ShortTitle{Transforming XRB Astrophysics with SKA+VLBI}

\author[1\ast]{Tao An \orcidlink{0000-0003-4341-0029} }
\ShortName{T. An et al.} 
\author[2,3\ast]{Ailing Wang \orcidlink{0000-0002-7351-5801}}
\author[4\ast]{James C. A. Miller-Jones\orcidlink{0000-0003-3124-2814}}
\author[5]{Valeriu Tudose\orcidlink{0000-0001-5317-220X}}
\author[6,7]{Pikky~Atri\orcidlink{0000-0001-8125-5619}}
\author[8]{Lang~Cui \orcidlink{0000-0003-0721-5509}}
\author[3]{Hua~Feng \orcidlink{0000-0001-7584-6236}}
\author[9]{Benito~Marcote\orcidlink{0000-0001-9814-2354}}
\author[10,11]{Sara E.~Motta\orcidlink{0000-0002-6154-5843}}

\affiliation[1]{Department of Astronomy, University of Science and Technology of China, Hefei, Anhui 230026, China}
\emailAdd{antao2008@ustc.edu.cn}
\affiliation[2]{State Key Laboratory of Particle Astrophysics, Institute of High Energy Physics, Chinese Academy of Sciences, Beijing 100049, China}
\affiliation[3]{Yunnan Observatories, Chinese Academy of Sciences, 396
Yangfangwang, Kunming, 650216, China}
\emailAdd{ailing.wang.wal@gmail.com}
\affiliation[4]{International Centre for Radio Astronomy Research -- Curtin University, GPO Box U1987, Perth, WA 6845, Australia}
\emailAdd{james.miller-jones@curtin.edu.au}
\affiliation[5]{Institute of Space Science - INFLPR Subsidiary, Atomistilor 409, Magurele, Ilfov, RO-077125, Romania}
\emailAdd{tudose@spacescience.ro}  
\affiliation[6]{ASTRON, Netherlands Institute for Radio Astronomy, Oude Hoogeveensedijk 4, 7991 PD Dwingeloo, The Netherlands} 
\affiliation[7]{Department of Astrophysics/IMAPP, Radboud University, P.O. Box 9010, 6500 GL, Nijmegen, The Netherlands}
\emailAdd{atri@astron.nl}
\affiliation[8]{Xinjiang Astronomical Observatory, Chinese Academy of Sciences, 150 Science 1-Street, Urumqi, 830011, China}
\emailAdd{cuilang@xao.ac.cn}
\emailAdd{hfeng@ihep.ac.cn}
\affiliation[9]{Joint Institute for VLBI ERIC, Oude Hoogeveensedijk 4, 7991~PD Dwingeloo, The Netherlands.}
\emailAdd{marcote@jive.eu}
\affiliation[10]{INAF Osservatorio Astronomico di Brera, Via E. Bianchi 46, I-23807 Merate, Italy}
\affiliation[11]{Astrophysics, Department of Physics, University of Oxford, Keble Road, Oxford OX1 3RH, UK}
\emailAdd{sara.motta@inaf.it}
\affiliation[\ast]{Chapter Corresponding Author}

\abstract{X-ray binaries (XRBs) are unique laboratories where accretion, jets, strong gravity and magnetic fields can be probed on humanly tractable timescales. The phased SKA-Mid operating as a single, ultra-sensitive Very Long Baseline Interferometry (VLBI) element will transform radio studies of XRBs primarily through its time-domain capabilities: substantially improved sensitivity on VLBI baselines that include SKA-Mid in Bands 2 and 5, together with connected-element SKA-Mid imaging extending down to 0.35--1 GHz (Band 1), microarcsecond-precision astrometry for bright systems, high-fidelity polarimetry for the most strongly polarized sources, and rapid target-of-opportunity response. In synergy with global VLBI networks, SKA+VLBI will track the evolution of compact ejecta and compact jets on astronomical unit (AU) scales, measure frequency-dependent core shifts to infer magnetic field strengths and gradients, resolve disk--jet coupling during state transitions in real time, and determine precise distances and natal kicks via parallaxes and proper motions. Joint campaigns with future X-ray and optical telescopes will enable strictly simultaneous, multi-band constraints on accretion--ejection physics and on jet composition. We outline a quantitative program for \aastar\ and AA4, including cadenced, multi-frequency VLBI ``movies'' of jets over the first days of outbursts, an XRB astrometric census, and a core-shift survey, and we provide representative detection rates, magnetic field measurements and distance accuracies. These outcomes will set the microphysical foundation for jet physics across the mass scale from stellar-mass black holes and neutron stars to active galactic nuclei, and will establish SKA+VLBI as the definitive facility for time-domain, high-resolution XRB astrophysics.
}

\begin{document}
\maketitle

\newcommand{\actaa}{Acta Astron.} 
\newcommand{\araa}{ARA\&A} 
\newcommand{\aar}{A\&ARv} 
\newcommand{\aapr}{A\&ARv} 
\newcommand{\ab}{Astrobiol.} 
\newcommand{\aj}{AJ} 
\newcommand{\apj}{ApJ} 
\newcommand{\apjl}{ApJL} 
\newcommand{\apjs}{ApJSS} 
\newcommand{\ao}{Appl. Opt.} 
\newcommand{\apss}{Astro. \& Space Sci.} 
\newcommand{\aap}{A\&A} 
\newcommand{\aaps}{A\&AS.} 
\newcommand{\baas}{Bull. Am. Astron. Soc.} 
\newcommand{\caa}{Chinese A\&A} 
\newcommand{\cjaa}{Chinese J. A\&A} 
\newcommand{\cqg}{Class. Quantum Gravity} 
\newcommand{\gal}{Galaxies} 
\newcommand{\gca}{Geo. Cosmo. Acta} 
\newcommand{\icarus}{Icarus} 
\newcommand{\jcap}{JCAP} 
\newcommand{\jgr}{J. Geophys. Res.} 
\newcommand{\jgrp}{J. Geophys. Res. Planets} 
\newcommand{\jqsrt}{J. Quant. Spectrosc. Radiat. Transf.} 
\newcommand{\memsai}{Mem. SAIt} 
\newcommand{\mnras}{MNRAS} 
\newcommand{\nat}{Nature} 
\newcommand{\nastro}{Nat. Astron.} 
\newcommand{\ncomms}{Nat. Commun.} 
\newcommand{\nphys}{Nat. Phys.} 
\newcommand{\na}{New Astron.} 
\newcommand{\nar}{New Astron. Rev.} 
\newcommand{\physrep}{Phys. Rep.} 
\newcommand{\pra}{Phys. Rev. A} 
\newcommand{\prb}{Phys. Rev. B} 
\newcommand{\prc}{Phys. Rev. C} 
\newcommand{\prd}{Phys. Rev. D} 
\newcommand{\pre}{Phys. Rev. E} 
\newcommand{\prx}{Phys. Rev. X} 
\newcommand{\prl}{Phys. Rev. Let.} 
\newcommand{\psj}{Planet. Sci. J.} 
\newcommand{\planss}{Planet. Space Sci.} 
\newcommand{\pnas}{Proc. Natl Acad. Sci. USA} 
\newcommand{\procspie}{Proc. SPIE} 
\newcommand{\pasa}{PASA} 
\newcommand{\pasj}{PASJ} 
\newcommand{\pasp}{PASP} 
\newcommand{\rmxaa}{RMXAA} 
\newcommand{\sci}{Science} 
\newcommand{\sciadv}{Sci. Adv.} 
\newcommand{\solphys}{Sol. Phys.} 
\newcommand{\sovast}{Soviet Ast.} 
\newcommand{\ssr}{Space Sci. Rev.} 
\newcommand{\uni}{Universe} 

\section{Introduction}

X-ray binaries (XRBs) are interacting binaries in which a compact object (a black hole, neutron star, or white dwarf) accretes from a companion; in this chapter we focus on black-hole and neutron-star systems, where dramatic state changes in the accretion flow drive powerful, relativistic outflows. These jets are best studied in the radio band, and their causal connection to the accretion flow is a key question in high-energy astrophysics \citep[e.g.,][]{Fender2004, Remillard2006, Done2007}. High-time-resolution X-ray spectroscopy and timing studies determine the inflow geometry, while radio interferometry unveils the outflow. Closing this inflow--outflow feedback loop requires not only \textbf{high angular resolution and sensitivity}, but in particular \textbf{high cadence and accurate polarimetry}, all of which must be delivered on the hours-to-days timescales over which XRBs evolve. 

The Square Kilometre Array Observatory (SKAO) will operate two telescopes: SKA-Low, 50--350~MHz; and SKA-Mid, 0.35--15.4~GHz. In the staged program, \aastar\ delivers early operations with 307 stations (SKA-Low) and 144 dishes (SKA-Mid), ramping to AA4 with $512$ stations and 197 dishes for the Design Baseline. Bands~1,~2,~5a,~5b are deployed first on SKA-Mid. Phasing SKA-Mid provides a tied-array beam that can augment existing global VLBI networks, yielding substantial improvements in baseline sensitivity and tens-of-microarcsecond  ($\mu$as) astrometry precision for bright XRB cores, while preserving the SKA imaging capability for the arcsecond-to-arcminute structure that trace jet--environment interactions. These properties, together with rapid target-of-opportunity (ToO) response, enable time-resolved VLBI ``movies'' of jet formation and quenching, a regime only glimpsed so far in a few remarkable outbursts (e.g., GRS~1915+105, MAXI~J1820+070; \citealt{Hjellming1995, Tingay1995, Jeffrey2016, Miller-Jones2019}).

The scientific opportunity of SKA+VLBI is threefold: (i) Jet microphysics and launch. Rather than directly resolving the true jet-launching region (which lies at $\lesssim 10^{3} - 10^{4}$ gravitational radii and is far below the $\gtrsim 10^{6} R_{g}$ angular resolution achievable for Galactic XRBs), SKA+VLBI will constrain magnetohydrodynamic (MHD) acceleration and magnetic topology by tracking the kinematics, spectral evolution and core shifts of compact ejecta on AU scales and by mapping how these change with accretion state \citep{BlandfordZnajek1977, McKinneyNarayan2007, Tchekhovskoy2011}. With resolutions of $\theta_{\rm beam} \approx 0.5\,\mathrm{mas}$ at 15\,GHz on intercontinental baselines, and positional uncertainties of $\sigma_{\rm pos}\approx \theta_{\rm beam}/(2\,\mathrm{SNR})$, where SNR is signal-to-noise ratio, tens of $\mu$as relative astrometry becomes routine for mJy-level cores, which is precisely the regime SKA+VLBI opens. Polarization and Faraday tomography in the brightest, most strongly polarized systems will then diagnose ordered versus turbulent magnetic fields through time-resolved polarization fractions, electric-vector position angles, and global Faraday rotation signatures, subject to the SNR limitations discussed below. (ii) Disk--jet coupling in the time domain is probed as the radio core brightens, flares, depolarizes, and quenches on hour--day timescales during X-ray state transitions. Coordinated SKA+VLBI plus X-ray coverage measures radio lags, spectral evolution, and structural changes with the cadence required to disentangle causality \citep{Fender2009, Russell2019, Wood2025}; while failed transitions and recurrent mini-outbursts probe threshold physics in jet launching. (iii) Precision astrometry and population physics are enabled by parallaxes and proper motions, which yield distances, luminosities, and natal kicks. Anchoring radio positions in a well defined inertial frame such as the International Celestial Reference Frame (ICRF; \citealt{2020A&A...644A.159C}) provides the absolute astrometry needed to reconstruct Galactic orbits, measure natal kicks and test compact-object formation channels, with consistency checks against the \textit{Gaia} optical frame where suitable counterparts exist \citep[e.g., ][]{Mirabel2001, Atri2019, Atri2020, Miller-Jones2021, Cui2025}.

This chapter is structured as follows. Section 2 consolidates technical capabilities focused on SKA‑Mid and their implications for VLBI observations of XRBs. Section 3 presents the core science applications structured around jet microphysics and time-domain evolution, core-shift magnetometry, astrometric natal-kick studies, and jet–ISM feedback. Section 4 discusses multi‑wavelength synergies with current and forthcoming facilities. Section 5 outlines observational strategies, including survey design, trigger logic, calibration, and pipelines, explicitly distinguishing AA*-achievable programmes from those requiring the full AA4 baseline. Section 6 summarizes the key outcomes.

\section{Technical Capabilities of SKA+VLBI for XRB Observations}

SKA-Mid Bands 2 (0.95--1.76\,GHz) and 5 (4.6--15.4\,GHz) provide the primary SKA+VLBI frequency coverage for opacity studies, core--shift measurements, and Faraday rotation/depolarization, while Band 1 (0.35--1.05\,GHz) adds lower-frequency leverage through connected-element SKA imaging and, where suitable partner stations have compatible UHF receivers, potential low-frequency VLBI extensions \citep{braun2019}. Phased-array operation delivers a tied beam to global VLBI networks (EVN, AVN, LBA), improving sensitivity on all baselines that include SKA \citep{Paragi2015aska}.
On intercontinental baselines, $\theta \simeq \lambda/B_{\max}$ implies $\approx 1~\mathrm{mas}$ at 8~GHz, $\approx 0.5$~mas at 15~GHz. For a typical XRB core with $S_\nu \gtrsim$ 0.5--5~mJy at 5--15~GHz during outburst, detectability can be estimated via the single-baseline thermal noise,
$\sigma_{\rm bl} \simeq \frac{\sqrt{{\rm SEFD}_1{\rm SEFD}_2}}{\eta_s\sqrt{2\,\Delta\nu\,t}}$,
where ${\rm SEFD}_{1,2}$ is the system equivalent flux density (in unit of Jy) of telescopes \#1 and \#2, $\eta_s$ is correlation efficiency (dimensionless), $\nu$ the bandwidth (Hz) and $t$ integration time (s);  
so that for a phased SKA-Mid element with ${\rm SEFD}_1\sim 2$--3~Jy paired with a 100~m-class antenna with ${\rm SEFD}_2\sim 20$~Jy one expects $\sigma_{\rm bl}\approx 30$--40~$\mu$Jy in $t=60$~s for $\Delta\nu=256$~MHz (and $\approx 10$--15~$\mu$Jy in $t=10$~min, scaling as $(\Delta\nu\,t)^{-1/2}$). This corresponds to ${\rm SNR}\approx 15$  for a 0.5~mJy core in 1 minute, and ${\rm SNR}\gtrsim 50$ in $\lesssim 1$--3~min for 1--5~mJy cores. In multi-baseline imaging, the naturally weighted map \textit{rms} for a global array including SKA-Mid reaches a few~$\mu$Jy~beam$^{-1}$ (e.g., $\sim$3~$\mu$Jy~beam$^{-1}$) in $\sim$1~h and scales approximately as $t^{-1/2}$ (so an 8~h track approaches $\sim$1~$\mu$Jy~beam$^{-1}$), depending on participating antennas and processed bandwidth. Astrometric precision scales as
$ \sigma_{\mathrm{pos}} \simeq \theta_{\mathrm{beam}}/(2\,\mathrm{SNR}) \lesssim (5\text{--}30)~\mu\mathrm{as} $
for SNR = 50--100 at 15 GHz, to which one must add a similar contribution from residual calibration systematics. For AA* we therefore adopt a conservative single-epoch astrometric error budget of $\sigma_\pi \lesssim 10-20~\mu\mathrm{as}$ for well calibrated, mJy-level targets, improving further in AA4 thanks to higher SNR and denser calibrator grids. This is sufficient for parallaxes with $\sigma_\pi \lesssim 10~\mu\mathrm{as}$  in multi-epoch programmes. 

Full-Stokes VLBI with mixed-feed calibration \citep{Marti-Vidal2016} and SKA’s polarization purity will enable RM synthesis \citep{Brentjens2005} across Band 5 for the brightest systems. However, typical linear polarization fractions $p$ in XRB cores and ejecta are only of order 1--3 per cent \citep[e.g.,][]{Brocksopp2013, Fender2002a, Fender2002b, Russell2014CygX1Pol}, so that detecting linear polarization at (for example) $5\sigma$ requires Stokes~I detections at ${\rm SNR}_I \gtrsim 5/p \approx 170$--500 for $p=0.03$--0.01. Polarimetric imaging at VLBI resolution will therefore be practical only for a subset of the best targets and will often yield statistical trends rather than detailed per-source RM maps.

Rapid ToO triggers from X--ray and optical monitors (e.g., Einstein Probe, Swift, MAXI, Rubin) are supported by SKA’s staged delivery timelines; the SKAO roadmap anticipates public science--verification data in 2027 and early operations from 2029, matching the start of \aastar\ exploitation for XRB time--domain VLBI campaigns.

 \aastar\ provides 144 dishes (SKA-Mid) and 307 stations (SKA-Low); AA4 grows to 197 and 512, respectively. Bands 1, 2, 5a, and 5b will be installed initially on Mid, with Bands 3--4 to follow when funded. In the remainder of this chapter, we explicitly note \aastar-achievable outcomes (e.g., bright outburst VLBI, parallax of the brightest subset) vs.\ AA4 expansions (e.g., fainter and quiescent targets, denser monitoring, larger samples).

\section{Scientific Applications}

Fundamental questions remain unsettled: How are jets launched and collimated? What governs disk--jet coupling? What role do magnetic fields play? SKA+VLBI will resolve these outstanding problems through $\mu$Jy-level imaging sensitivity, $10\mu$as-level astrometry, and rapid target-of-opportunity (ToO) response, enabling real-time observation of jet dynamics, quantitative mapping of accretion flows, and population-scale constraints on compact object formation across the stellar-mass regime. For an SKA perspective on accreting compact-object binaries, please refer to \citet{Beri01.2026.SKA}.

\subsection{Resolving the Inner Workings of Relativistic Jets}

The formation and early evolution of relativistic jets from X-ray binaries remain among the most compelling puzzles in high-energy astrophysics \citep{Mirabel1994, Mirabel1999,Fender2004}. 
SKA+VLBI’s key advance is not direct imaging of the event-horizon-scale jet-launching region, which remains inaccessible for Galactic XRBs, but rather the ability to follow the evolution of compact jets and discrete ejecta on AU scales with high cadence and high astrometric precision. These time-resolved observations, combined with broadband spectra and polarimetry, provide powerful indirect constraints on jet microphysics and launch.

\begin{figure}
    \centering
    \includegraphics[width=0.5\linewidth]{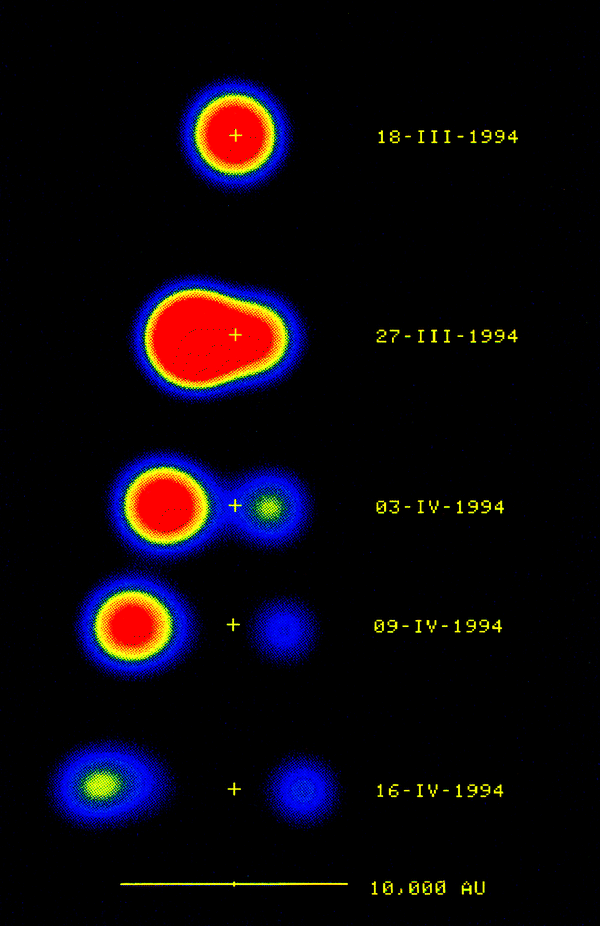}
    \caption{Benchmark for jet kinematics applied to Soft X-ray Transients (SXTs). Time-sequence radio images of GRS 1915+105 from 18 March to 16 April 1994 show the birth of compact knots from the core (cross) and their separation with apparent superluminal motion ($\beta_{\rm app} > 1$); scale bar: 10,000 AU. Adapted from \citet{Mirabel1994}. SKA-Mid phased for global VLBI will extend this kinematic diagnostic to faint, short-lived ejecta in SXTs, where jets are often thought to be ``quenched''. $\mu$Jy sensitivity and sub-mas imaging enable (i) hour--day--month cadence proper-motion tracking of weak knots, depending on their temporal evolution, (ii) back-extrapolated ejection timestamps to test whether jet launch precedes or follows the soft-state transition at AU scales, (iii) measurements of deceleration, opening angle, and core-shift to infer magnetic fields and kinetic power. Thus, the Mirabel-Rodríguez-type analysis becomes a population tool to quantify jet energetics and duty cycles across the entire XRB class.
}
    \label{fig:Mirabel1994}
\end{figure}

The transformative power of SKA+VLBI lies in its capacity to track jet evolution with unprecedented temporal resolution while retaining sub-mas angular resolution. 
Time sequences of VLBI images during the first days of an outburst will reveal the birth, acceleration and expansion of individual knots, building on the pioneering work on systems such as GRS 1915+105 and MAXI J1820+070 \citep{Mirabel1994, Dhawan2007, Bright2020, Miller-Jones2019, Wood2025}. Enhanced sensitivity will enable continuous monitoring of the trajectories of much fainter ejecta, yielding precise measurements of apparent superluminal motions, accelerations and deviations from ballistic motion \citep{Mirabel1994,Russell2019, Wood2025}, which encode the interaction between jet plasma and ambient medium \citep{Fender2025}. A canonical example of this kinematic approach is shown in Figure~\ref{fig:Mirabel1994}.

Although the innermost jet-launching and collimation zone will not be spatially resolved, SKA+VLBI can still probe it indirectly. Frequency-dependent core positions, spectral indices, and time lags between radio and X-ray signatures carry information about where along the flow different optical depths and particle acceleration sites occur \citep{Markoff2001, Markoff2005, Heinz2004}. By measuring how the core position and spectrum respond to changes in accretion state, we can test MHD jet models that tie jet power and magnetization to the inner disk and black-hole spin \citep{BlandfordZnajek1977, BlandfordPayne1982, Tchekhovskoy2011, Blandford2019, BaanAn2025,  Pathak01.2026.SKA}. Counter-jet detections in the brightest systems will break degeneracies between intrinsic speed and inclination \citep{Fender1999, Miller-Jones2021}, improving estimates of jet kinetic power.

Spectral-index mapping along resolved ejecta, especially when combining multi-frequency VLBI and connected-element SKA images, will reveal where and how particles are accelerated and cooled \citep{Markoff2001,Markoff2005,Fender2004,Miller-Jones2012}. 
Rather than resolving transverse structures such as spine–sheath configurations or systematic transverse RM gradients, which are generally beyond realistic angular resolution and sensitivity for XRBs, SKA+VLBI will primarily constrain longitudinal gradients and global asymmetries. Any hints of transverse structure are likely to arise first in the brightest, most nearby systems and would require stacking and careful visibility-domain modelling.

The rapid evolutionary timescales of XRB jets mean that phenomena that would take centuries to millennia to unfold in AGN are compressed into weeks to months \citep{FenderBelloni2004,Markoff2015}. 
SKA+VLBI therefore offers a unique opportunity to test jet physics across a wide range of conditions, with the lessons feeding directly into our understanding of relativistic outflows from supermassive black holes \citep{HeinzSunyaev2003, Gallo2003, Merloni2003, Falcke2004, Kording2006}.

\subsection{Disk-jet coupling through rapid state transitions}

\begin{figure}
    \centering
    \includegraphics[width=0.8\linewidth]{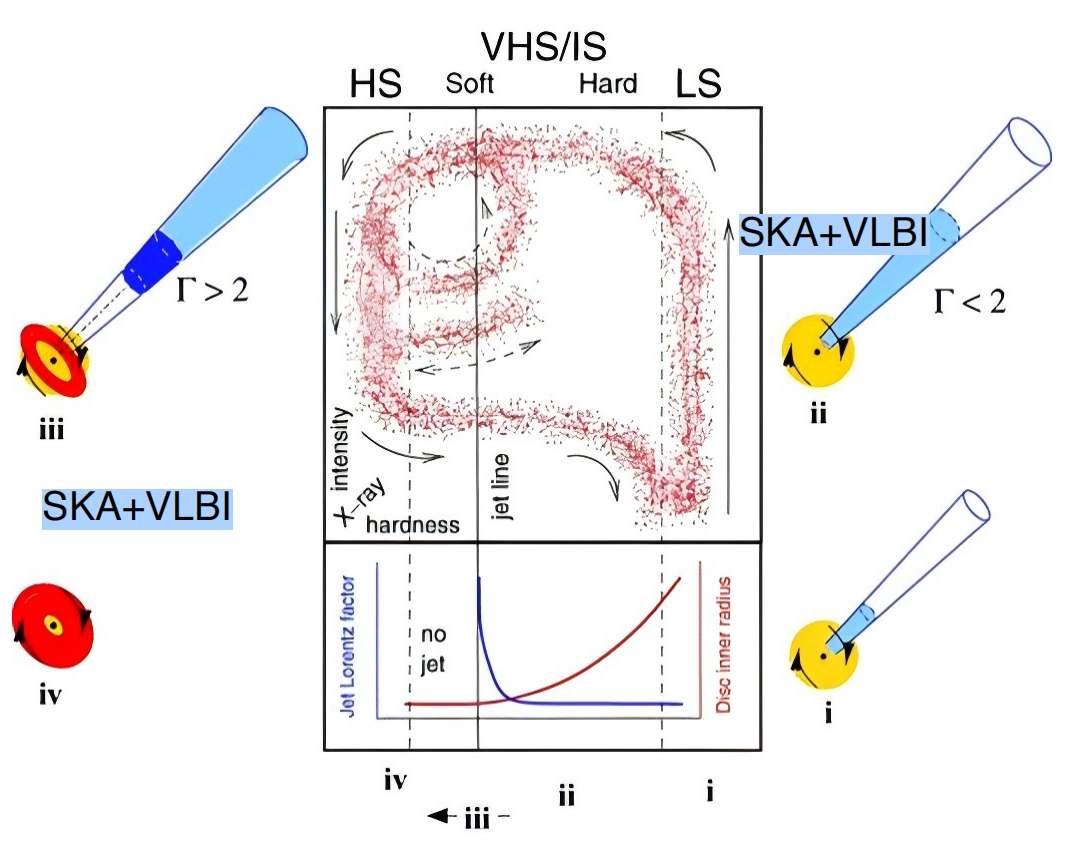}
    \caption{Fender–Belloni–Gallo schematic of X-ray binary state transitions and associated jet modes \citep{Fender2004}, annotated to illustrate the unique diagnostic power of SKA + VLBI. In the \textit{hard state}, SKA+VLBI will resolve and astrometrically monitor the compact, self-absorbed core, enabling ejection timing measurements to timestamp jet launch relative to the X-ray spectral transition at AU scales. During the \textit{hard-to-soft transition}, triggered VLBI imaging can follow knot emergence and measure apparent speeds (bulk Lorentz factors) and opening angles. In the \textit{soft state}, the SKA's $\mu$Jy-level continuum imaging sensitivity permits the definitive test of whether radio ``quenching'' represents a total jet shutdown or a merely extreme suppression of the outflow. Together, these capabilities transform the phenomenological XRB state diagram into a quantitative, time-resolved framework linking accretion-state evolution, jet launching, and magnetic-field topology across multiple scales.}
    \label{fig:Fender}
\end{figure}

The close connection between accretion flows and relativistic jets represents one of the most fundamental aspects of black hole physics \citep{Fender2004,Done2007}. X-ray binaries undergo dramatic state transitions where accretion flow structure changes within days to weeks, accompanied by launching or quenching of relativistic jets \citep{Remillard2006,Belloni2011}. SKA+VLBI's  combination of sensitivity and rapid response capabilities enable us to probe the causal connection between accretion flow changes and jet evolution. The canonical Fender--Belloni--Gallo state diagram is shown in Figure~\ref{fig:Fender}.

During state transitions, SKA+VLBI will allow dense monitoring of jet formation and destruction. Previous observations show that hard-to-soft state transitions are often accompanied by discrete ejection events \citep{Fender2009,Miller-Jones2012, Russell2014MAXIJ1836}, but the exact sequence leading to these ejections remains poorly understood. With SKA+VLBI, we can begin intensive campaigns within hours of an X-ray trigger, obtaining a small number of long ($\geq $8 h) tracks over the first few days to follow ejection events from their earliest stages, resolving their motion and expansion and relating these to changes in the radio core and X-ray spectra.

Of particular interest is the  process of jet quenching during soft state transitions \citep{Fender2004, Fender2009,Coriat2011}. SKA+VLBI cannot see the innermost engine switching off but can resolve the compact radio core down to AU scales, enabling us to determine whether the radio ``quenching'' corresponds to a complete disappearance of compact synchrotron emission or to an extreme suppression below current sensitivity limits. By comparing SKA+VLBI measurements with concurrent lower-resolution SKA imaging we can test for the presence of any residual diffuse core emission. Correlating these radio signatures with X-ray timing properties \citep{Kalemci2016} will provide crucial insights into how accretion flow geometry affects jet production.

Rapid cadence and high sensitivity will also improve our understanding of jets in intermediate states, where accretion flows may be rapidly reconfiguring \citep{Fender2014}. These observations will reveal whether jet properties change smoothly or abruptly  during state transitions, addressing long-standing questions about jet launching and stability \citep{Markoff2005}. Simultaneous tracking of both the compact core and ejected blobs will show how the jet base responds to accretion flow changes while previous ejections propagate outward.

Failed state transitions, where systems start to soften but then return to the hard state, offer a particularly clean probe of threshold physics \citep{Corbel2013}. 
SKA+VLBI imaging of such events will show how close a system can get to quenching the jet while still maintaining a radio core, and what changes occur in jet structure and speed. Combining these with X-ray spectral and timing diagnostics will allow us to map out the regions of parameter space where jets can and cannot be sustained.

In this way, XRBs observed with SKA+VLBI will provide a unique time-resolved view of accretion--ejection coupling that can be scaled up to supermassive black holes \citep{Kording2006,Merloni2003} helping to unify the physical picture of how black holes convert infalling matter into powerful relativistic outflows.

\subsection{Astrometric Studies} 

How do stellar mass black holes form and do they all form the same way? Do their formation channels resemble those of neutron stars? What factors influence how a black hole is formed: progenitor mass, metallicity, binary interactions, or a more complex combination of factors? Nearly a decade after the discovery of black hole mergers from gravitational waves that has widened the mass distribution of known stellar mass black holes, we are still seeking answers to these questions. Simulations present a complex, non-monotonic relationship among initial stellar mass, BH mass, and formation pathway \citep{Sukhbold2016,Burrows2025}. Black hole XRBs provide one of the few laboratories where theories of stellar mass black hole birth can be tested. 

The two most common theoretical scenarios for creating a BH are: (a) direct collapse, whereby the massive progenitor star collapses directly into a BH with very little or no mass ejection \citep[e.g.,][]{Adams2017}; or (b) delayed formation in a supernova, in which fallback of material ejected during the explosion onto the proto-neutron star creates a BH \citep{Fryer2006}. Unlike neutron stars, only one BHXRB candidate has a confirmed, direct association with a supernova remnant \citep[SS433;][]{Dubner1998}, making it difficult to determine whether black hole birth involves successful SN explosions.

Natal kick is the additional velocity imparted to a compact object during a successful supernova explosion, and is a probe to differentiate between direct collapse and supernova explosion birth scenarios for black holes. 
Neutron star kicks have been studied through the proper motions of pulsars \citep[e.g.][]{Hobbs2005}, and the space velocities of neutron stars in binary systems \citep[e.g.][]{O'Doherty2023}. This has shown evidence for a low-kick population (required for binary survival, and to explain the prevalence of X-ray binaries in globular clusters) and a much higher-kick population, primarily observed in young pulsars \citep[e.g.][]{Verbunt2017}.
Efforts are underway to estimate kick velocities and infer the birth mechanism of BHXRBs \citep{Atri2019,Nagarajan2025}, suggesting that most such systems have low kicks. However, these studies suffer from small sample statistics with only a nominal 15-20 sources available for study as compared to $>200$ neutron star systems.

Other than helping in successful discrimination between the two theoretical birth models for BHs, natal kicks are key parameters to map the distribution and motion of Galactic BHXRBs, which are necessary for understanding their environments. Natal kicks are crucial inputs to estimate BH retention fractions in globular clusters \citep{Della2024}, the formation rate of binary BH systems that could be the progenitors of GW events \citep{Fragione2021} and BH-BH merger sites \citep{Kelley2010}. The black hole spin-orbital plane misalignment is also highly sensitive to natal kicks \citep{Wysocki2018,Poutanen2022}, and are essential links to unearth the origin of BH binaries \citep{Naoz2025}. Current BH natal kick distribution assumptions for the above simulations and for stellar binary evolution codes are made by comparing them to NS natal kicks \citep[e.g.][]{Mandel2020}, or involve flat distributions up to some cutoff \citep[e.g.,][]{Sippel2013}, underscoring the importance of obtaining accurate natal kick constraints for the BHXRB population.

Natal kick estimations require the measurement of the full three-dimensional space velocities of XRBs, reconstruction of their Galactocentric orbits back to the time of their formation and estimating the kick velocity that might have sent the XRB into its current orbit. These velocities inform detailed evolutionary analysis of each system to determine the natal kicks that are in turn able to explain the current observed properties of the system \citep{Willems2005,Fragos2009}. To infer the three-dimensional motion of BHXRBs, proper motion and distances of BHXRBs are folded in with the systemic radial velocity. For sources that are too optically faint for Gaia (including the bulk of XRBs), VLBI astrometry is, by far, the most precise means to obtain proper motions and parallaxes of XRBs \citep{Reid2011, Miller-Jones2014, 2009ApJ...706L.230M}. XRBs are distributed across the Galaxy at distances from 1 kpc to possibly 25 kpc \citep[e.g., BW Cir;][]{Plotkin2021} and precise distances are one of the most important parameters to convert any observed parameter into a physical property of the system. The current milliarsecond resolution VLBI arrays limit reliable distance measurements of these usually faint radio sources (< 1 mJy) to under 5kpc, biasing our studies to only the closest systems. 

The huge leap in sensitivity, combined with microarcsecond-precision astrometry from phased SKA-Mid on VLBI baselines \citep{Rioja01.2026.SKA}, will open an unprecedented window into the faintest, slowest and farthest XRBs, enabling a far more complete study of the Galactic XRB population. The increased sensitivity will enable proper motion and parallax measurements of nearby XRBs in quiescence, circumventing the current necessity of systems being in outburst for them to be bright enough to have detectable radio emission during the hard state. SKA+VLBI will make fainter calibrator sources accessible, allowing for in-beam and multiview phase calibration \citep{Rioja2017}, that will drastically reduce the calibrator-target throw based systematics, compounding sensitivity based improvement in astrometric accuracy. The increased accuracy of position measurement at each instance will decrease the required time baseline needed to significantly measure a proper motion of a few mas yr$^{-1}$. Since many XRBs go into short lived outbursts, this will be revolutionary for increasing the sample size of systems with measured proper motions. 
By bringing us into the microarcsecond astrometry era for sub-mJy sources, SKA+VLBI will expand our astrometric reach to the entire Galactic XRB population. SKA+VLBI will finally provide statistically robust natal kick measurements for the XRB population, transforming our understanding of compact object formation and their evolution in binaries. Related SKA-enabled population studies of Galactic binary systems are discussed in \citet{Marcote01.2026.SKA} in this Volume, and for an SKA-VLBI perspective on compact objects, please refer to
\citet{Lin01.2026.SKA} in this Volume.

Microarcsecond-level astrometry will also allow us to measure orbital motion in a subset of high-mass or nearby systems \citep[e.g.][]{Miller-Jones2021}, tracking the motion of compact objects during their binary orbits. Precise measurements of orbital sizes and inclinations, coupled with radial-velocity measurements that probe the donor star orbit, will provide some of the first model-independent estimates of the component masses \citep[e.g.][]{Andrews2019}. 

\subsection{Mapping Magnetic Field Evolution through Polarimetric Imaging}

The role of magnetic fields in jet launching, collimation, and evolution represents a fundamental puzzle in high-energy astrophysics \citep{Blandford2019}. SKA+VLBI's  polarimetric capabilities will provide new constraints on magnetic-field structure in XRB jets, especially in the brightest systems
\citep{Marti-Vidal2016}. 
High-sensitivity polarimetric imaging will reveal the magnetic field configuration in compact jet cores and in resolved ejecta where polarization is detectable \citep{Brocksopp2013, Fender2002a, Fender2002b}. Detection of systematic Faraday rotation measure (RM) gradients across jet width will provide direct evidence for helical magnetic fields, providing a key prediction of magnetohydrodynamic jet launching models \citep{BlandfordZnajek1977,McKinneyNarayan2007}. However, such transverse gradients require both high angular resolution and high SNR in Stokes Q and U, and will therefore likely not be measurable even in the brightest nearby, strongly polarized systems. SKA+VLBI will instead constrain the global RM and polarization fraction as a function of time and frequency.

Time-resolved polarimetric observations during state transitions \citep{Fender2004} will reveal how accretion flow changes affect jet magnetic field structure. For example, changes in polarization angle and depolarization can signal the emergence of new components or changes in the relative contribution of different jet regions. High sensitivity will permit detection of polarized emission from both bright discrete ejecta, showing how magnetic fields evolve as knots expand and interact with the ambient medium, although many fainter ejecta will remain below the detection threshold in polarization even when they are well detected in total intensity.

Broad frequency coverage will probe frequency-dependent polarization properties of XRB jets, revealing internal Faraday rotation and depolarization and constraining the mix of ordered and turbulent magnetic fields \citep{Brocksopp2013, Russell2014CygX1Pol}. 
Combining VLBI polarization with lower-resolution SKA measurements will allow us to separate small-scale internal Faraday effects from larger-scale screens, leading to a more complete picture of XRB jet energetics and composition.

\subsection{Measuring frequency-dependent core shifts to constrain magnetic field strength and gradient}

XRB compact radio jets exhibit a ``core'' (the apparent bright jet base) that is a self-absorbed synchrotron region whose position depends on observing frequency \citep{BlandfordKonigl1979,Heinz2004}. At higher frequencies, lower optical depth places the observed core closer to the black hole; at lower frequencies, the $\tau \approx 1$ surface lies further downstream. This frequency-dependent core position shift is a well-established phenomenon in relativistic jets \citep{BlandfordKonigl1979, Lobanov1998}, first clearly detected in an XRB jet for SS433 \citep{Paragi1999} and subsequently observed in other microquasars such as Cyg X-3 \citep{Tudose2007,Egron2021} and GRS 1915+105 \citep{Dhawan2000}. For typical Galactic XRBs at a few kpc, the expected shifts between 5 and 15 GHz are of order 0.01–0.1 mas, that is, well matched to the astrometric precision achievable with SKA+VLBI.

Because synchrotron optical depth depends on physical parameters (magnetic field strength, particle density, etc.), changes in jet conditions (e.g., increased flow density during flares) cause core position to move outward \citep{Sharma2022,Paragi1999, Heinz2004}. Measuring core shift as a function of frequency thus provides a powerful diagnostic of innermost jet structure \citep{Lobanov1998,Sokolovsky2011,Sharma2022}. In particular, it enables probing the magnetic field ($B$) in the jet and how $B$ varies with distance from the black hole with profound implications for understanding jet launching and collimation mechanisms in XRBs, especially in comparison to AGN jets.

In practice, multi-frequency astrometry measures the position of the synchrotron ``core'' (the $\tau\simeq 1$ surface) as a function of observing frequency. In a conical, partially self-absorbed jet this follows $r_{\rm core}(\nu)\propto \nu^{-1/k_r}$, so fitting the measured offsets $\Delta r$ between frequency pairs yields a core-shift parameter (often written $\Omega_{r\nu}$) and the index $k_r$ (e.g., \citealt{BlandfordKonigl1979,Lobanov1998,Sokolovsky2011}). With a distance estimate and modest assumptions about geometry and Doppler factor, $\Omega_{r\nu}$ can be converted into an estimate of the magnetic field strength at a fiducial distance (e.g., 1 parsec from the central BH, commonly quoted as $B_{1\,\rm pc}$ in AGN work) and then extrapolated inward via $B(r)\propto r^{-m}$, where $m\simeq 1$ is expected for a predominantly toroidal magnetic field in an approximately conical flow. The same framework also provides an estimate of the de-projected location of the jet apex relative to the observed VLBI core, enabling direct comparison to X-ray timing constraints on the inner accretion flow.

Until now, empirical core shift studies in XRBs have been sparse \citep[e.g.][]{Prabu2023}, largely due to extreme angular resolution and precision requirements. Expected shifts are on the order of milliarcseconds to microarcseconds for Galactic sources \citep[e.g.][]{Zdziarski2012}. SKA+VLBI will revolutionize this area. By incorporating the SKA into global VLBI networks, significantly enhanced sensitivity on baselines including SKA-Mid and microarcsecond-precision astrometry for bright sources will become more readily achievable. SKA-Mid Band 5 (4.6--15.4 GHz) covers exactly the high-frequency range ideal for core--shift measurements.  Precise alignment of reference frames between different frequencies is also crucial. This task will be greatly facilitated by the dense grid of fainter calibrators that SKA+VLBI makes available. With SKA+VLBI, astrometric accuracy of a few $\mu$as is achievable for sources of order 1 mJy, enabling routine core shift measurements in a large sample of Galactic black hole XRBs for the first time.

The scientific payoff is significant: spatial structure mapping of the radio-emitting zone and direct inference of magnetic field strength and gradient along the jet. These measurements provide insight into jet composition and energetics, offering a new handle on accretion--ejection physics. By comparing core-shift-derived $B$ fields across different accretion states, we can investigate how jet magnetization changes when sources transition from steady jets (hard X-ray state) to weak 
or extinguished jets (soft state). Similar techniques applied to AGN have recently linked jet magnetic profiles to accretion parameters (even black hole spin), so applying them to more rapidly evolving XRBs can test such ideas in a new regime.

The high-frequency capability of SKA-Mid, especially Bands 5a and 5b, corresponds to $<10$ AU spatial scales for typical sources at a few kpc, sufficient to resolve shifts in core position expected between 5 GHz and 15 GHz. The plan is to observe microquasar jets at multiple frequencies across Band 5, ideally simultaneously or within very short time intervals to minimize intrinsic source variability effects. High cadence is also important, especially during outbursts, jet conditions can evolve on days-to-weeks timescales, so we envision scheduling SKA+VLBI session series to track core shift as sources brighten and fade.

In summary, SKA+VLBI frequency-dependent core shift measurements will yield: (1) magnetic field strength at the jet base, (2) magnetic field gradients along the inner jet, (3) constraints on particle distribution and magnetic dominance, (4) time-resolved diagnostics of jet magnetization evolution during outbursts, and (5) improved jet base location estimates. Combined with simultaneous X-ray (and possibly $\gamma$-ray/optical) monitoring, these measurements will test models of accretion--ejection coupling, probe correlations between jet magnetic field and X-ray luminosity.

\subsection{Probing jet-environment interactions across multiple scales}

\begin{figure}
    \centering
    \includegraphics[width=0.5\linewidth]{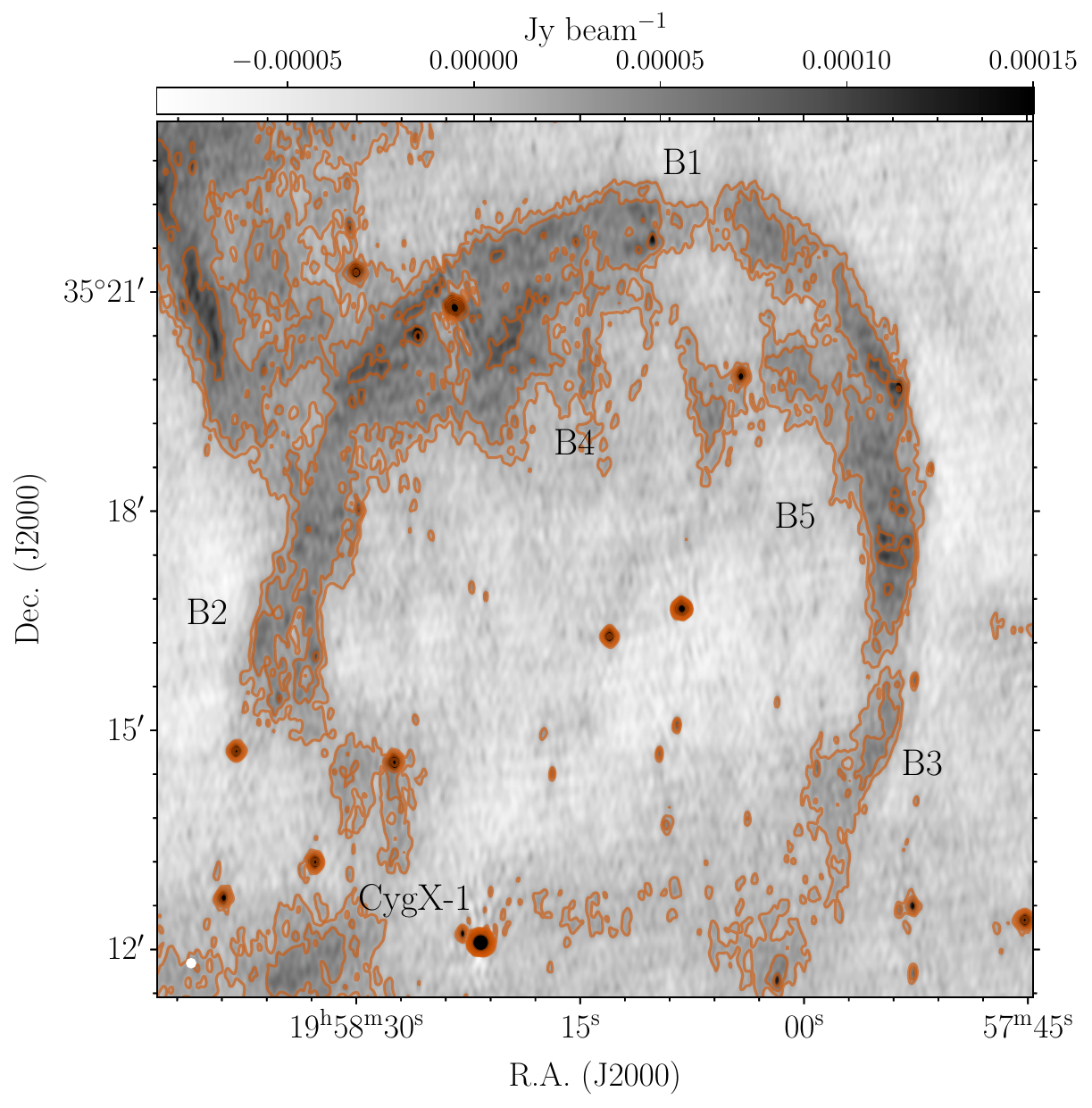}
    \caption{MeerKAT imaging has revealed a shell and hotspot system around Cygnus~X-1, including a compact core and a $\sim$20~pc ring seen at 1.28 and 2.64~GHz \citep{Atri2025}. The shell (the U-shaped curve denoted by B2-B1-B3) and the substructures (B4, B5) within the shell are consistent with powerful jets impacting the interstellar medium \citep{Gallo2005}. Similar jet-inflated bubbles have recently been discovered around another microquasar (e.g., GRS~1915+105; \citep{Motta2025}). SKA+VLBI is not intended to image these large structures directly; instead it constrains the early, compact phase by measuring jet proper motions, opening angles, spectral evolution (including core shift), and polarization, providing the kinetic-power input required to model the subsequent bubble evolution. The large-scale shocked plasma and relic electrons are then best mapped by SKA-Low and SKA-Mid connected-element imaging over 50~MHz--15~GHz.} 
    \label{fig:CygX1Bubble}
\end{figure}

The way XRB jets deposit energy and momentum into their surroundings is most cleanly diagnosed on parsec scales, where jet-inflated bubbles, bow shocks and synchrotron lobes can be resolved \citep{Heinz2002, Gallo2005, Corbel2005, Russell2007, Bright2020, Atri2025, Mariani2025, Motta2025}. These phenomena are best probed with connected-element imaging by VLA and MeerKAT, and future SKA-Mid and SKA-Low. However, interpreting such large-scale structures in physical terms requires a handle on the time-averaged kinetic power, duty cycle and orientation history of the inner jets that feed them, which is precisely where SKA+VLBI adds unique value. An illustrative example of a jet-blown bubble and ring system, as already revealed by MeerKAT, is shown in Figure~\ref{fig:CygX1Bubble}.

In this picture, SKA+VLBI is used only during the early, compact phase of an outburst, when individual ejecta are still unresolved or marginally resolved on intercontinental baselines. Phased SKA-Mid on VLBI baselines will measure proper motions, apparent speeds and opening angles of knots on AU scales, and will track how their brightness and spectra evolve over the first few days before they expand and resolve out \citep{Russell2019, Bright2020}. From these data one may infer the instantaneous kinetic energy and momentum flux carried by the jets, together with the duty cycle of powerful ejection episodes over the lifetime of the source. Once the ejecta are resolved out at VLBI resolution, the same systems are followed at arcsecond scales with SKA-Mid and SKA-Low, which map the jet-inflated bubbles, bow shocks and steep-spectrum relic emission on 0.1--100 pc scales \citep{Pakull2010, Corbel2012}.

SKA-Low will be particularly powerful for detecting faint, old bubbles and outer bow shocks through their steep-spectrum emission, allowing spectral-aging and Mach-number diagnostics. SKA-Mid Bands 1–2 will add broadband polarization and RM synthesis to trace shock-compressed magnetic fields and the ambient ISM geometry. In this regime SKA+VLBI no longer provides direct images of the extended structures, but the AU-scale information obtained earlier remains crucial: by combining the jet speeds, inclination angles and ejection frequencies measured with SKA+VLBI with the energetics and morphologies of the bubbles seen by SKA, we can perform a true energy and momentum calorimetry, testing whether the large-scale structures are consistent with the integrated jet power and duty cycle inferred from the inner jets.

High-precision SKA+VLBI astrometry also links the inner-jet axis and the binary’s space motion to the larger-scale morphology. For example, comparing the VLBI-measured jet direction with the axis of the SKA-resolved bow shock or ring can test whether the jet direction has remained stable over many outbursts or has precessed, and whether asymmetric structures are shaped primarily by jet variability or by the motion of the system through the ISM. This multi-scale approach, in which SKA+VLBI constrains the ``instantaneous engine'' and SKA imaging constrains the ``time-integrated footprint'', turns spectacular morphologies into quantitative laboratories for jet feedback.

\subsection{Ultraluminous X-ray sources as extragalactic laboratories}

Ultraluminous X-ray sources (ULXs) provide an important extension of XRB phenomenology into the super-Eddington regime, and at least a subset can be explained as stellar-mass black holes or neutron stars accreting at extreme rates \citep[e.g.,][]{Kaaret2017, Bachetti2014}. Several nearby ULXs show compact radio counterparts, transient flares, or collimated outflows detected with high-resolution radio interferometry (e.g., \citealt{Middleton2013M31,Cseh2014HoIIX1,Cseh2015HoIIX1}), and some inflate large radio bubbles on 10--100~pc scales (e.g., \citealt{Lang2007NGC5408,Soria2010S26}).
 SKA-Mid imaging will map and characterize the diffuse bubbles and bow shocks, quantifying the time-averaged mechanical power, while SKA+VLBI can provide sub-mas localization and brightness-temperature constraints on any compact core emission, separating genuine ULX jets from background AGN. 
For the nearest systems ($D\lesssim 5$--10~Mpc), mas scales correspond to $\sim 10^{-2}$~pc, allowing structural-evolution constraints during bright radio flares \citep[e.g.,][]{Middleton2013M31,Cseh2015HoIIX1} and a direct comparison between instantaneous (compact-core) and time-integrated (bubble) energy output \citep[e.g.,][]{Kaaret2003SciNGC5408,Lang2007NGC5408,Soria2010S26}. Including ULXs in the SKA+VLBI programme therefore provides a natural extragalactic complement to the Galactic transient sample and tests whether the jet and outflow physics inferred in sub-Eddington XRBs extends into the super-Eddington regime \citep[e.g.,][]{Kaaret2017,Fabrika2004}.

\subsection{Unveiling the Nature of Soft X-ray Transients}

The Einstein Probe mission \citep{Yuan2018einstein,Yuan2025} is now uncovering a diverse population of soft X-ray transients, including events plausibly associated with accreting compact objects in the Galaxy and nearby galaxies. Recent examples that may be related to the XRB population include the black hole XRB candidate EP~J182730.0$-$095633 \citep{Cheng2025EPJ182730}, a rare Be--white-dwarf XRB candidate EP~J005245.1$-$722843 in the Small Magellanic Cloud \citep{Marino2025EPJ005245}, and intermediate-duration EP transients whose nature remains under debate \citep{Zhang2025EP240408a}. 
These sources are characterized by low-luminosity soft X-ray spectra in the 0.3--10~keV band, rapid variability on minute--hour timescales, and complex spectral evolution, challenging existing XRB phenomenology at the faint and fast end. SKA+VLBI provides a unique opportunity to unravel their physical nature and establish their connection to the classical XRB framework \citep{Fender2004,Fender2009}. 

Einstein Probe provides the critical trigger and early-time context for this programme by discovering and localising soft X-ray transients on timescales that match the rapid evolution of XRB outflows. The key SKA+VLBI contribution is therefore concentrated in the first days after an EP trigger, when compact radio cores and newly launched ejecta remain detectable on intercontinental baselines; after this window, the science naturally transitions to connected-element SKA imaging of expanding, lower-surface-brightness emission. We therefore envisage an EP-driven trigger logic that prioritises rapid SKA snapshots for classification and flux-density assessment, followed by a small number of long ($\gtrsim 8$ h) SKA+VLBI tracks during the first 1–3 days for the best candidates, enabling AU-scale kinematics and astrometric registration. Broader multi-wavelength coordination, including complementary roles of X-ray monitors and optical surveys, is described in Section \ref{sec:multiwavelength}.

\section{Multi-wavelength Synergy} \label{sec:multiwavelength}

The SKA \citep{braun2019} is expected to advance radio studies of XRBs through its improved sensitivity, survey capabilities, and timing resolution. By incorporating the SKA into global VLBI networks, mas-resolution imaging of AU-scale jets and ejecta, together with microarcsecond-precision astrometry of compact radio cores, will become routine for the best targets. In addition, ejecta can be tracked with higher cadence during outbursts, enabling direct measurements of proper motions and expansion. Also, astrometry at microarcsecond precision will provide accurate distance estimates even for faint sources.

Historically, X-ray observations have been central to advancing our understanding of XRBs, providing key insights into accretion states, variability, and spectral transitions. The integration of SKA with VLBI networks will complement the capabilities of wide-field X-ray monitors and pointed observatories that drive XRB triggering and state classification, including MAXI, Swift, NICER, Insight-HXMT, and Einstein Probe, as well as future missions such as Athena and eXTP. In particular, the Einstein Probe provides a complementary discovery channel for XRB-like fast transients, delivering rapid, well-characterized soft X-ray triggers and early-time spectral and temporal evolution that can prioritise SKA follow-up. The combination of SKA+VLBI timing and spatial resolution will enable high-fidelity investigations of X-ray and radio correlations and rapid variability, refining models of jet launching, collimation, and their coupling to accretion flows.

In addition to enhancing studies of outbursting systems, SKA+VLBI will allow to probe the size scale of quiescent jets \citep{Heinz2006, Miller-Jones2008, Prabu2023}. These low-luminosity regimes are particularly important for constraining radiatively inefficient accretion flows. 

On the optical/IR side, SKA+VLBI will play a major role in complementing astrometric measurements from missions such as \textit{Gaia}. 
Anchoring radio positions in a well defined inertial frame such as the International Celestial Reference Frame (ICRF) provides the absolute astrometry needed to reconstruct Galactic orbits, measure natal kicks and test compact-object formation channels, with consistency checks against the Gaia optical frame where suitable counterparts exist.
By providing parallaxes and proper motions for XRBs that are too optically faint or heavily extincted for Gaia, SKA+VLBI will extend the census of accurately distance-calibrated systems into regions of the Galactic plane and bulge that are inaccessible in the optical, while still allowing consistency checks with the Gaia optical frame where suitable counterparts exist.

The synergy between optical transient surveys and SKA+VLBI will also open new avenues for studying jet formation and early outflow evolution in XRBs. Detections of new XRB transients by the Vera C. Rubin Observatory (VRO) can act as triggers for rapid SKA+VLBI follow-up, enabling observations of jet activity close to the compact object at the earliest stages of an outburst. Such multiwavelength efforts will allow precise measurements of jet launching timescales, collimation, and interaction with the surrounding environment, providing important constraints for theoretical models of inflow--outflow coupling.

\section{Observational Strategies}

The SKA+VLBI observational programme for X-ray binaries represents a paradigm shift in time-domain, high-resolution astrophysics. Its goal is to transform sporadic snapshots of jet evolution into carefully targeted, high-cadence sequences during the brief intervals when VLBI is most informative, and to build statistically meaningful samples of astrometric measurements and core-shift detections.
This strategy is built around three synergistic pillars: (i) time-resolved imaging of relativistic jets during outbursts, (ii) precision astrometry to construct a comprehensive census of XRB distances and space motions, and (iii) multi-frequency campaigns to measure magnetic field structures via core-shift diagnostics and polarimetric mapping. These programmes span both the early operational stage (\aastar) and the full design baseline (AA4), ensuring transformative science from first light through full deployment.

The time-resolved jet imaging programme constitutes the cornerstone of SKA+VLBI's attack on the jet formation problem. For each suitable outburst, triggered by alerts from X-ray or optical facilities, SKA-Mid could participate in VLBI at Bands 5a and 5b with a small number of long tracks 
during the critical first two weeks post-outburst, when the jets are still compact enough to be well detected on intercontinental baselines. This temporal baseline is chosen to capture both the launch and early expansion of discrete ejecta,  while avoiding inefficient oversampling once the jets have expanded and become resolved out at VLBI scales. 
In \aastar, campaigns will target the brightest transients ($S_\nu \gtrsim 1$ mJy), establishing the methodology and building the first systematic sample. AA4 will enable broader coverage, including fainter systems and somewhat denser early-time sampling for the very best events. Visibility-domain modelling techniques, such as those recently applied to Swift J1727.8‑1613 and similar systems, e.g. \citep{Wood2025}, will be routinely employed to extract kinematics and expansion rates even when components are only marginally resolved in the image plane.

Astrometric observations form the backbone of a long-term XRB parallax and proper-motion census. 
Given that there are fewer than one hundred known black-hole and black-hole-candidate XRBs in the Galaxy and that many are too faint in quiescence for VLBI, a realistic initial goal is to track of on the order of 20–30 radio-bright systems over 2–4 years, with 8–10 epochs per source strategically spaced to optimize parallax factors and decouple proper motion. This sample can include both black-hole and neutron-star XRBs. The VLBI reference frame will be tied to nearby calibrators using in-beam or multi-view phase referencing when possible \citep{Rioja2017}. 
For sources at approximately 5 kpc, expected positional precision of $\sigma \lesssim$ 10--20 $\mu$as will yield distance errors below a few percent and three-dimensional space velocities for systems with known radial velocities. These measurements are vital for constraining the compact object's birth environment, natal kick velocity, and binary evolutionary history.

A third pillar is the core shift and magnetometry of outflowing systems (CoSMOS) programme. This  will employ quasi-simultaneous VLBI observations at four or more frequency points across Bands 2, 5a, and 5b, measuring spatial offsets of the synchrotron core as a function of $\nu$ to yield direct constraints on magnetic field strength $B(r)$ and its radial profile along the jet. In \aastar, CoSMOS will target a carefully selected subset of $\approx 10-15$ well-studied, bright XRBs with compact, steady jets in the hard state. AA4 will extend this to $\approx 20$ sources and to somewhat fainter cores. By comparing core shifts and inferred magnetic profiles across different accretion states and compact-object types, CoSMOS will test whether jet magnetization evolves systematically during outbursts and whether black-hole and neutron-star jets occupy the same parameter space.

SKA's fast-response observing mode will support tiered responses: rapid snapshot imaging within 24 hours of trigger alerts (at SKA resolution) to identify promising radio-loud transients (\citealt{GemmaAnderson01.2026.SKA} in this Volume); intensive SKA+VLBI campaigns with long tracks during the first few days after ejection for those sources; and lower-cadence follow-up with SKA and pathfinder arrays during the decline, once the jets have expanded beyond the optimal VLBI scale. All programmes will integrate swift quality assurance and quick-look imaging pipelines, with data rates of several Gbps per station anticipated. Where appropriate, subarrays will be used to balance sensitivity, (u,v) coverage and survey speed. In total, this strategy exploits SKA's dual role as both a precision interferometer and a fast-cadence transient machine, while remaining realistic about the small number of bright XRBs and the short time window during which VLBI is most effective. It will establish SKA+VLBI as the definitive facility for time-domain, high-resolution studies of accretion and ejection physics in Galactic XRBs.

\section{Summary}

X-ray binaries are laboratories for studying accretion, jets, strong gravity, and magnetic fields on humanly tractable timescales. The phased SKA-Mid, operating as an ultra-sensitive VLBI element, will transform radio studies of XRBs by delivering substantially improved sensitivity in SKA imaging over 0.35--15.4~GHz and on VLBI baselines primarily over 1--15.4~GHz, microarcsecond astrometry, high-fidelity polarimetry, and rapid target-of-opportunity response. SKA+VLBI will track compact jets and ejecta on AU scales, measure frequency-dependent core shifts to infer magnetic-field strengths and gradients, and follow disk–jet coupling during state transitions in real time. Through a combination of time-resolved jet imaging, long-baseline astrometric campaigns, and multi-frequency core-shift studies, SKA+VLBI will deliver precise distances and natal kicks via parallaxes and proper motions, quantitative constraints on jet kinematics and energetics, and measurements of magnetic-field structure in and around the jet.
Coordinated campaigns with X-ray and optical telescopes, and future missions will enable strictly simultaneous, multi-band constraints on accretion--ejection physics and jet composition. We outline a quantitative programme spanning \aastar\ to AA4, including cadenced multi-frequency VLBI ``movies'' of jets during the first days of outbursts, an XRB astrometric census targeting $\approx 20-30$ systems, and the core-shift survey. These outcomes will establish the microphysical foundation for jet physics across the mass scale from stellar-mass black holes and neutron stars to active galactic nuclei, positioning SKA+VLBI as the definitive facility for time-domain, high-resolution XRB astrophysics.

\section*{Acknowledgment} 
We thank the reviewer for the constructive comments. 
TA acknowledges the Shanghai Oriental Talent Project.
MPT and JM acknowledge financial support from the Severo Ochoa grant CEX2021-001131-S and from the Spanish grant PID2023-147883NB-C21, funded by MCIU/AEI/ 10.13039/501100011033, as well as support through ERDF/EU. MPT and JMW also acknowledge the service and support from the Spanish Prototype of an SRC (SPSRC), funded by the Spanish Ministry of Science, Innovation and Universities, by the Regional Government of Andalusia, by the European Regional Development Funds and by the European Union NextGenerationEU/PRTR.
VT acknowledges support from the Romanian Ministry of Research, Innovation and Digitalization through the Romanian National Core Program LAPLAS VII – contract no. 30N/2023.
PA is supported by the WISE fellowship program, which is financed by NWO.
BM acknowledges financial support from the State Agency for Research of the Spanish Ministry of Science and Innovation, and FEDER, UE, under grant PID2022-136828NB-C41/MICIU/AEI/10.13039/501100011033 and via an Advanced Grant from the European Research Council (ERC) under the European Union’s Horizon 2020 research and innovation programme (`EuroFlash’; Grant agreement No. 101098079).
AI tools were used solely to improve the language and readability of the manuscript.

\bibliographystyle{abbrvnat-maxbibnames4}
\bibliography{AASKAII_ID50_VLBI_XRB} 

\end{document}